\documentstyle[12pt]{article}

\makeatletter
\@addtoreset{equation}{section}
\makeatother

\newcommand{\be}{\begin{equation}}
\newcommand{\ee}{\end{equation}}
\newcommand{\bee}{\begin{eqnarray}}
\newcommand{\eee}{\end{eqnarray}}

\newcommand{\ga}{\alpha}
\newcommand{\gb}{\beta}

\newcommand{\nn}{\nonumber}
\newcommand{\half}{\frac{1}{2}}

\newcommand{\sgm}{\mbox{$\sigma_{-}$}}

\newcommand{\sgp}{\mbox{$\sigma_{+}$}}
\newcommand{\sgpg}{\mbox{$\tilde \sigma_{+}$}}
\newcommand{\sgmg}{\mbox{$\tilde \sigma_{-}$}}

\newcommand{\ah}[1]{\mbox{$a_#1$}}
\newcommand{\ac}[1]{\mbox{$a_#1^+$}}
\newcommand{\cet}{\rangle}
\newcommand{\f}{\mbox{$\wedge$}}
\newcommand{\al}{\left (}
\newcommand{\ar}{\right )}

\newcommand{\ra}{\rangle}
\newcommand{\la}{\langle}
\newcommand{\D}{\mbox{${\cal D}$}}
\newcommand{\p}{\mbox{$P_\bot$}}

\newcommand\un{{\underline{n}}}
\newcommand\um{{\underline{m}}}

\newcommand{\ti}{\tilde}
\newcommand{\Dg}{\mbox{$\ti {\cal D}$}}

\begin{document}

\thispagestyle{empty}

\begin{flushright}
\vspace{1mm}
FIAN/TD/08--00\\
{March 2000}\\
hep-th/xxxyyzz\\
\end{flushright}

\vspace{1cm}

\begin{center}
{\large\bf
Scalar Field in Any Dimension {}From the
 Higher Spin Gauge Theory Perspective  }\\
\vglue 1  true cm
\vspace{1cm}
{\bf O.V.~Shaynkman
\footnote{e-mail: shayn@lpi.ru}
  and M.A.~Vasiliev }
\footnote{e-mail: vasiliev@lpi.ru}  \\
\vspace{1cm}

I.E.Tamm Department of Theoretical Physics, Lebedev Physical
Institute,\\
Leninsky prospect 53, 117924, Moscow, Russia
\vspace{1.5cm}
\end{center}

\begin{abstract}
We formulate the equations of motion of a free
scalar field in the flat and $AdS$ space of an  arbitrary dimension
in the form of some ``higher spin" covariant constancy conditions.
Klein-Gordon equation is interpreted as a non-trivial
cohomology of a certain ``\sgm-complex".
The action principle for a scalar field is formulated in terms of
the ``higher-spin" covariant derivatives for an arbitrary mass in
$AdS_d$ and for a non-zero mass in the flat space.
The constructed action is shown to be
equivalent to the standard first-order Klein-Gordon action at the
quadratic level but becomes different at the interaction level
because of the presence of an infinite set of auxiliary fields which
do not contribute at the free level. The example of Yang-Mills
current interaction is considered in some detail.
It is shown in particular
how the proposed action generates the pseudolocally exact form of
the matter currents in $AdS_d$.

\end{abstract}

\newpage

\section{Introduction}\label{intro}

Mikhail Vladimirovich
Saveliev was a brilliant scientist who made a fundamental
contribution to  the theory of integrable systems.
Misha was always open to a scientific discussion and ready to
share his knowledge to colleagues.
Working mainly on two-dimensional integrable systems,
during last years Saveliev was interested
in the application of his ideas to the study
of higher-dimensional relativistic supersymmetric models \cite{MS}
which, he believed, have a chance to be solvable in one or another
sense. Now it is an open problem to clarify
to which extend his expectations were true.

The aim of this contribution is to reformulate the simplest
relativistic model of a scalar field in any dimension in a way
inspired by the theory of higher spin gauge fields. So far,
the higher spin gauge theory has been developed beyond the linearized
level mainly for  $d\leq 4$
(see \cite{rev1,rev} and references therein).
{}From these particular cases  it is known that
higher spin gauge theories are based on
appropriate infinite-dimensional symmetries,
higher spin symmetries. All massless fields with spins $s\geq 1$
are gauge fields. In the framework of the formalism developed in
 \cite{more}, the $d=4$
dynamical equations have a form of some zero-curvature
and covariant constancy conditions supplemented with certain constraints.
It remains to be analyzed whether this is a signal of some sort of
integrability of the 4d  higher spin models.
The reformulation of the scalar field dynamics suggested in this
paper is useful as a starting point towards yet unknown
infinite-dimensional higher spin symmetries
in any dimension. Also, it sheds some light on the specificities
of the higher spin interactions.

The action principle compatible with the higher spin gauge symmetries
and general covariance to the lowest nontrivial order in interaction
was formulated in $AdS_4$ \cite{FV1},
thus solving the problem of introducing consistent gravitational
interactions of higher spin gauge fields to the  cubic order in
interactions. The cubic action of \cite{FV1} was however known to be
incomplete requiring some further modification at the higher orders in
interactions. Indeed, it is well known that the consistency
of some interactions at the cubic level does not guarantee that the
theory can be consistently extended beyond the cubic order.
(For example, at the cubic level one
can consider any number N of spin 3/2 gravitinos interacting with
gravitons but only for $N\leq 8$ it is possible to
proceed beyond the cubic level with very specific sets of fields,
the supergravitational supermultiplets, carrying spins $s\leq 2$.
The true spectrum of fields can only be fixed  at the quartic level.
The higher spin interactions beyond the cubic order were studied in
\cite{Ann,more} at the level of equations of motion. {}From these
results and also from the analysis of the unitary lowest weight
representations of the higher spin algebras (i.e. higher-spin
multiplets) in \cite{KV1} it is known that the full spectra of spins
in the complete higher spin theories contain lower spin
fields with spins 1 and 1/2 and 0.
One of the aims of this paper is to construct a
spin-0 action in the form analogous to the spin $s>1$ actions used
in \cite{FV1} as a step towards a complete higher spin action.
As argued in \cite{FV1} unbroken higher spin symmetries require AdS
geometry rather than the flat one.
We therefore consider the problem both in the flat and  $AdS$
background.

Infinite-dimensional higher spin symmetries mix higher derivatives
of all orders  of the dynamical fields. To make these symmetries
manifest it is useful to reformulate dynamics in terms of
appropriate  higher spin covariant derivatives $\D C^A (x)$.
The representation $C^A (x)$ of the higher spin symmetry is
infinite-dimensional containing infinitely many field components
(index $A$), most of which  express via the
higher derivatives of the dynamical fields by virtue of certain
constraints. We formulate the action principle
for a free scalar field in terms of the covariant derivative $\D C$
in $AdS_d$ for an arbitrary value of mass $m$ and
 in the flat space for $m\neq 0$.
At the free field level the constructed action is equivalent
to the standard first-order Klein-Gordon action because
higher components of the representation $C^A$ do not contribute to the
action at the quadratic level thus guaranteeing it to be of the normal
order in derivatives. However, the proposed action
differs from the standard one at the interaction level.
In particular, in this framework  minimal Yang-Mills interaction
of a scalar field turns out to be combined with
some additional interactions to the Yang-Mills field strength
containing infinite series in higher derivatives of the
scalar field with the coefficients proportional to
negative powers of the parameter of mass or the cosmological constant.
An immediate  consequence of our formulation is that it shows
how the Yang-Mills current built from the scalar matter can be compensated
by a pseudolocal field redefinition, which result provides a
generalization to an arbitrary dimension  of the observation
made for 3d higher spin models \cite{PV2} that conserved currents are
pseudolocally exact in $AdS_3$. Also let us mention  some parallelism
between our results and the description of the  consistent interaction of
massive higher spin fields with gravity
in the recent papers \cite{per_,per},
which
requires infinite expansions in negative powers in the parameter of mass
at the action level. Needless to say that this picture is reminiscent of
the $\ga^\prime$ expansion for massive modes in the string theory.

The paper is organized as follows. In the Section 2, we
formulate the equations of motion for a scalar field of an
arbitrary mass in  the flat space in the ``unfolded" form of
certain covariant constancy conditions. These results are
generalized to $AdS_d$ in the Section 3.
In the Section 4,  the Klein-Gordon equations are interpreted
in terms of the \sgm-cohomology.
In the Section 5, we derive the scalar field
action formulated in terms of the
``higher spin'' covariant derivatives of the Sections 2 and 3.
The  Fock space notation are introduced in the Section 6.
Specificities of the Yang-Mills current
interaction of the scalar fields described by the action of
the Section 5 are
discussed in the Section 7.  In the Section 8 we generalize the
covariant constancy conditions of the sections 2 and 3 to the
 off-mass-shell system. The Section 9 contains some conclusions.

\section{``Unfolded" Formulation of the Scalar Field Equations in the
        Flat Space}

To describe  a free massless scalar field $c(x)$ in $d$ dimensions,
let us introduce a set of traceless symmetric tensors
of all ranks $C(x)=(c(x),\ldots,c^{n(k)}(x),\ldots)$,
\be
\label{tr0}
\eta_{nn} c^{ n(k+2)} = 0\,,
\ee
where $m\,,n \,,\ldots = 0,\ldots ,d-1$ and $\eta^{nm}$ is the  mostly minus
flat metric
$ \eta^{nm} = (1,-1,\ldots ,-1 )$. In this paper we use the conventions
of \cite{V2} convenient for the component analysis
of complicated tensor structures: upper and lower indices denoted by the
same letter should be first symmetrized (separately) and then the maximal
possible number of lower and upper indices should be contracted;
the number of indices can be indicated in brackets by writing e.g.
$n(k)$ instead of writing $k$ times the index $n$.
Underlined Latin indices are used for differential forms and vector fields
in d-dimensional space-time with coordinates $x^\un$,
\be
\um\,,\un \,,\ldots = 0,\ldots ,d-1\,,\qquad
\partial_\un = \frac{\partial}{\partial x^\un}\,,\qquad
d=dx^\un \partial_\un \,.
\ee
Indices from the middle of the Latin Alphabet are fiber.

As shown in \cite{rev1} the
equations of motion of a free massless scalar field $c(x)$ in the flat
$d-$dimensional space-time can be reformulated
as the following infinite chain of equations
\be
\label{un0}
\partial_\un c_{n(k)} +e_\un{} {}^m c_{n(k)m}=0\,,
\ee
where $e_\un{} {}^m$ is the flat space vielbein.
Such a form of the dynamical equations expressing all the derivatives
in terms of the fields was called ``unfolded'' in \cite{unf}.
In the flat space
one can choose $e_\un{} {}^m=\delta_\un^m$ and  identify
underlined (base) and non-underlined (fiber) indices.

The first two equations in (\ref{un0}) read
\be
\label{1}
\partial_\un c =-c_\un\,,
\ee
\be
\label{2}
 \partial_\un c_\um=- c_{\um\un}\,.
\ee
Eq. (\ref{1}) tells us that
$c_\un$ is the first derivative of $c$.
Eq. (\ref{2}) implies that
$c_{\un\um}$ is the second derivative of $c$. However, because of the
tracelessness condition
\be
\label{trC}
c_n{}^n =0\,,
\ee
it imposes the Klein-Gordon equation
\be
\label{KG}
\Box c =0\,.
\ee
The rest equations in (\ref{un0}) express highest
tensors $c_{n(k)}$ in terms of the higher-order derivatives
\be
\label{hder}
c_{\un_1 \ldots \un_k}=(-1)^k \partial_{\un_1}\ldots\partial_{\un_k}c
\ee
imposing no further conditions on $c$. The tracelessness conditions
(\ref{tr0}) are all satisfied once the Klein-Gordon equation
(\ref{KG}) is true.

Let $T^p_k$ be a linear space of $p$-forms taking values in the
space of rank-$k$ totally symmetric traceless tensors.
In other words, a general element of  $T^p_k$ is
$c^{n(k)}=dx^{\um_1}\wedge \ldots\wedge dx^{\um_p}c_{\um_1 \ldots
\um_p}{}^{n(k)}$, with $ c_n{}^{n(k-1)}=0$.
Then $T^p = \sum_{k=0}^\infty \oplus T^p_k$
is the linear space of $p$-forms taking values in the
space of all totally symmetric traceless tensors.

Let us introduce the operator
\be
\sgm :\quad T^p_k\rightarrow T^{p+1}_{k-1}\,,
\label{sgmdiag}
\ee
\be
(\sgm C)^{n(k-1)}=
e_{m}\f c^{n(k-1)m}\,,\qquad \sigma_- (T_0^p)=0\,,
\label{formsgm}
\ee
where $e_{n}$ is the frame 1-form  $ e^n = dx^\un e_\un{}^n $.
(In the flat space we can set $e^{n}=dx^\un\delta_{\un}^n = dx^n$.)
The operator $\sigma_-$ has the following important properties
\be
\label{sigpr}
(\sigma_- )^2 =0 \,,\qquad
\sigma_- d + d \sigma_- =0\,.
\ee
The chain of equations (\ref{un0}) then takes the form
\be
\label{DC0}
(d+\sgm)C=0\,.
\ee
The compatibility condition
$(d+\sgm)^2=0$ of this system holds as
a consequence of (\ref{sigpr}).

Let us now address the question of the uniqueness of the equation
(\ref{DC0}) within the class of equations formulated in $T^p$
in terms of the exterior differential and the frame 1-form.
The only Lorentz covariant possibility is to write a chain of equations
\be
\label{cald0}
{\cal D} C=0 \,,
\label{chain}
\ee
\begin{equation}
{\cal D} = d+\sgm+ \sgp \,,
\label{fleq}
\end{equation}
with  \sgp : $T^p_k\rightarrow T^{p+1}_{k+1}$ being some
operator of the form
\be
(\sgp C)^{n(k+1)} = f(k)\p( e^{ n} c^{ n(k)})\,,
\label{formsgp}
\ee
where $f(k)$ are some unknown coefficients and
\p{} is the projector to the subspace of traceless tensors, i.e.
\be
(\sgp C)^{n(k+1)} = f(k)\Big (e^{ n} c^{ n(k)}
-\frac{k}{d+2(k-1)}\eta^{ n n}e_{m}
c^{ n(k-1) m}\Big )\,.
\label{formsgp1}
\end{equation}

The compatibility condition
\be
{\cal D}^2 =0
\ee
requires in addition to (\ref{sigpr})
\begin{eqnarray}
&\sgp d + d \sgp = 0\label{flsgp1}\,,\\
&\sgp\sgp=0   \label{flsgp2}\,,\\
&\{\sgp,\sgm\}=0\label{flsgp3}\,.&
\end{eqnarray}

The first two conditions are trivially satisfied while the third
imposes the following restrictions on the coefficients
$f(k)$:
\be
f(k)=\frac{(k+1)(d+2(k-1))}{k(d+2k)}f(k-1)
\label{flf}\,.
\ee
The generic solution of these equations is
\be
f(k)=-m^2 \frac{k+1}{2k+d}\,,
\ee
where $m^2$ is an arbitrary constant.

The first two equations in the chain (\ref{chain}) give
\be
\partial_\un c +e_\un{}{}^m c_m=0\,,\label{fleq1}
\ee
\be
\partial_\un c_n +e_\un{}{}^m c_{nm}-\frac{m^2}{d}e_{\un n}c=0\,.
\label{fleq2}
\ee

Contracting indices in the second equation and
substituting $c_n$ from (\ref{fleq1}) we obtain
\begin{equation}
(\Box+m^2)c=0.
\label{flcg}
\end{equation}
Therefore, the ambiguity in the coefficients $f(k)$ expresses the
ambiguity in the parameter of mass in the Klein-Gordon equation
reformulated in the form (\ref{cald0}). An equivalent formulation
for the massive scalar field in the flat space of an arbitrary
dimension was given in \cite{PV}. Note that
(\ref{hder}) has the same form for any value of $m^2$.

\section{``Unfolded"  Scalar Field Equations in $AdS_d$}

To generalize these results to $AdS_d$,
consider the gauge fields $A_\un{}^{MN}= -A_\un{}^{NM}$ for
the $AdS_d$ algebra $o(d-1,2)$, ($M,N=0,\ldots,d$).
Setting  $\omega_\un{}^{nm}=A_\un{}^{nm}$ and
$e_\un{}^n=\lambda^{-1}A_\un{}^{nd}$, where $\lambda$ is a constant, the
$o(d-1,2)$--- Yang-Mills strengths acquire the form
\begin{equation}
R_{\un\um}{}^{nm}=\partial_\un\omega_\um{}^{nm}+ \omega_\un{}^n{}_t
\omega_\um{}^{tm}-\lambda^2e_\un{}^ne_\um{}^m- (\un\leftrightarrow\um)\,,
\label{adscur}
\end{equation}
\begin{equation}
R_{\un\um}{}^n=\lambda\partial_\un e_\um{}^n+
\lambda\omega_\un{}^n{}_m e_\um{}^m-
(\un\leftrightarrow\um)\,.
\label{adstor}
\end{equation}
The fields $e_\un{}^n$ and $\omega_\un{}^{nm}$ are identified
with the vielbein and Lorentz connection.
Provided that ${\rm det} |e_\un{}^n| \neq 0$,
$\lambda^{-1}R_{\un\um}{}^n$ and $R_{\un\um}{}^{nm}$
identify, respectively, with the
 torsion tensor and Riemann curvature tensor (corrected by the
$\lambda-$ dependent ``cosmological term")
in the vielbein formulation of gravity. The equations
\be
\label{Rab}
R_{\un\um}{}^{nm}=0
\ee
 and
\be
\label{Ra}
R_{\un\um}{}^n=0
\ee
describe $d-$dimensional anti de-Sitter space of radius $\lambda^{-1}$.

The Lorentz covariant derivative is defined according to
\be
D_\un f^{nm\ldots}=
\partial_\un f^{nm\ldots}+\omega_\un{}^n{}_t  f^{tm\ldots}
+\omega_\un{}^m{}_t  f^{nt\ldots}+\cdots \,.
\ee
{}From (\ref{adscur})
and  (\ref{Rab}) it follows that in $AdS_d$
\begin{equation}
[D_\un,D_\um](f^{nm\ldots})=\lambda^2(e_\un{}^ne_{\um t} f^{tm\ldots} +
e_\un{}^me_{\um t}f^{nt\ldots}+\cdots )-(\un\leftrightarrow\um)\,.
\label{adscom1}
\end{equation}

To generalize the scalar field equation in the form
(\ref{cald0}) to  $AdS_d$ case, we set
\be
\label{adscovd}
{\cal D}= D+\sgm+\sgp
\ee
with $D= dx^\un D_\un$ and some new operator \sgp{}.
The compatibility condition $\D\D=0$ in the $AdS_d$ case
requires the same properties of \sgm{} and \sgp{} (\ref{sigpr}),
(\ref{flsgp1}), (\ref{flsgp2}), but modifies (\ref{flsgp3}) to
\be
\left(\{\sgp,\sgm\}C\right)^{n(k)}=-\left (DD(C)\right ) ^{n(k)}=
-k\lambda^2e^{n}\f e_m c^{n(k-1)m}
\label{sgp3}
\end{equation}
as a result of (\ref{adscom1}).
Taking again \sgm{} and  \sgp{} in the form (\ref{formsgm}) and
(\ref{formsgp}), respectively,
 (\ref{sgp3}) imposes the following conditions on $f(k)$
\be
f(k)=\frac{(k+1)(d+2(k-1))}{k(d+2k)}(f(k-1)+k\lambda^2)
\label{adsf}.
\ee
The generic solution of this equation is
\be
f(k)=\frac{k+1}{2k+d}(\lambda^2 k(k+d-1)-m^2 )\,,
\label{adsf_}
\ee
where $m^2$ is again an arbitrary parameter associated with the
solution of the homogeneous part of (\ref{adsf}).
Analogously to the flat case, we derive from the first
two equations of the chain (\ref{chain}) the massive Klein-Gordon
equation\footnote{\label{mlp}Let us note that there are several competing
definitions of a massless scalar field in $AdS_d$ with $m^2=m_0^2 \neq 0$.
For the  massless field  identified with the conformal scalar,
$m_0^2 =-\frac{\lambda^2}{4}d(d-2)$.}
\be
(\Box+m^2)c=0\,,\qquad \Box=\eta_{nm}D^nD^m
\label{adsKG}
\ee
and
\be
c^n=-D^n c\,,
\label{adsc1}
\ee
where $D^n=e^{\un n}D_\un$ (with
$e^\un{}_n e_\um{}^n = \delta^\un_\um$).
{}From the rest equations of the chain we obtain
a covariantized version of
(\ref{hder})
\be
c^{n_1\ldots n_k}=(-1)^k\p D^{n_1}\cdots D^{n_k}c\,.
\label{adshder}
\ee
Let us stress that the system (\ref{chain}) contains no
restrictions on $c$ beyond (\ref{adsKG}).
One way to prove this is to analyze cohomology
of the operator $\sigma_-$ (see section \ref{sec_cog}).

If some consistent
theory of fields of all spins possessing higher
spin gauge symmetries exists, the operator ${\cal D}$
should be interpreted as a result of linearization of
full nonlinear higher spin covariant derivatives near
the $AdS_d$ vacuum solution with the background $AdS_d$ gauge
fields being of zero order. Other way around, one can
take the form of the covariant derivative (\ref{adscovd})
as a starting point towards the full higher spin symmetry
and  its matter field representations. This strategy
was proved to be successful in \cite{FVA} for the
analysis of $d=4$ higher spin dynamics starting from
the appropriate  generalization of the covariant derivative
 (\ref{adscovd}) for the $d=4$ higher spin massless fields.
Also, the dynamical equations in the form (\ref{cald0})
is a good starting point towards
unfolded nonlinear higher spin equations (for more detail on
this point we refer the reader to \cite{rev1}).

\section{ \sgm{} Cohomology and Dynamical Equations}
\label{sec_cog}

An interesting feature of the proposed formulation is that
the equations of motion of a scalar field in the flat and
$AdS$ space admit a natural interpretation in terms of
the cohomology group of \sgm.
According to (\ref{sgmdiag}), \sgm{} increases a degree
of differential forms decreasing a number of tensor indices
and has the property $\sigma_-^2 = 0 $.
We show that the only nontrivial class
$H^1 (\sigma_- )$ of the first cohomology group of \sgm
\be
T\stackrel{\sgm}{\longrightarrow}\stackrel{H^1}{T^{1}}
\stackrel{\sgm}{\longrightarrow}T^2 \,,
\ee
 belongs to $T^1_1$ and describes the
left hand sides of the equations of motion for a scalar field.
The
constraints (\ref{hder}) or (\ref{adshder})  fix
particular representatives of the trivial cohomology
classes.

Consider the restriction $DC |_{T^1_0}$
of $DC$ to $T^1_0$ (i.e. the part in $DC$ that does not carry
vector indices).
By definition of \sgm,
\be
\label{start}
\sgm \left ( D C |_{T^1_0}\right )  =0\,.
\ee
The question is whether $D C |_{T^1_0} =
\sigma_-(y)$ with some $y^n\in T_1^0$. In components,
it is equivalent to $D_\un c =e_{\un n} y^n$.
Obviously, the solution of this equation exists
provided that the frame field $e_{\un{} n}$ is invertible.
Thus
 $H_0^1 =0$.
Since $y^n$ enters this equation just the same
way as $c^n$ enters (\ref{fleq1})
the fact that
$H_0^1 =0$ means that one can choose the field
$c^n$ in such a way that
\be
\label{X}
\left ((D+\sgm)C\right ) |_{T^1_0} =0\,.
\ee
Physically, this equation is interpreted as the constraint (\ref{fleq1})
expressing $c^n$ via the first derivatives of the physical field $c$.
Cohomologically, it fixes a representative of
the trivial cohomology class $c^n$ in terms of the derivatives
of the dynamical field $c$. Note that this condition
can be interpreted as a
constraint because \sgm{} does not contain space-time derivatives.

Since $(\sgp C)|_{T_0^1 }=0$, the condition (\ref{X}) is equivalent
to
\be
{\cal D} C|_{T_0^1}=0\,.
\ee
Now let us show that ${\cal D} C|_{T_{k+1}^1}$ is \sgm - closed if
\be
\label{hyp}
{\cal D} C|_{T_l^1}=0\qquad \mbox{for}\quad 0\leq l\leq k\,.
\ee
Indeed, since the operators $D$ and \sgp{} map $T_k$ to $T_l$ with
$l\geq k$  we obtain from (\ref{hyp}) that
\be
\left ((D+\sgp ){\cal D} C \right )|_{T_l^1}=0\qquad \mbox{for}\quad
l\leq k\,.
\ee
{}From ${\cal D}^2 = 0$ it follows then
that
\be
\left (\sgm {\cal D} C \right ) |_{T_l^1}=0\qquad \mbox{for}\quad l\leq k\,
\ee
and, therefore,
\be
\sgm ( {\cal D} C |_{T_{k+1}^1} )=0\,.
\ee
As a result, we conclude that
\be
\label{cc}
{\cal D} C |_{T_{k+1}^1} = \sgm (y_{k+2} ) +h_{k+1}\,,
\ee
where $\sgm (y_{k+2} ) $ describes some \sgm - exact part (i.e. a trivial
cohomology class) and $h_{k+1}$ is a representative of a
nontrivial cohomology $H^1$.
Obviously, the \sgm - exact part can be fixed to zero
by adjusting a field $c^{n(k+2)}$ on the
left hand side of (\ref{cc}). Therefore it is possible to impose
constraints
\be
\label{dc0}
{\cal D} C |_{T_{k+1}^1}=0
\ee
provided that the cohomology
group is zero. If it is different from zero, the equation (\ref{dc0})
imposes some differential equations on the bottom field $c(x)$.

Let us now find the cohomology group $H^1$.
A \sgm - closed  1-form $\hat c_{\un ,}{}^{n(k)}$ obeys
\be
e_{\un m}\hat c_{\um ,}{}^{n(k-1)m}-e_{\um m}
\hat c_{\un ,}{}^{n(k-1)m}=0\,.
\ee
Therefore
\be
e_{\un m}\hat c_{\um ,}{}^{n(k-1)m}=y_{\un\um ,}{}^{n(k-1)}\,,
\label{yfe}
\ee
where $y_{\un\um , }{}^{n(k-1)}$ is symmetric in
$\un$ and $\um\,$, and symmetric and traceless in $n(k-1)$.
Equivalently, using the fiber indices, we have
\be
\hat c_{m, n(k)}=y_{mn , n(k-1)}\,.
\ee
{}From this relation it follows that $y_{ml, n(k-1)}$ should be
totally  symmetric in all indices and traceless for
$k\geq 2$. This implies that $\hat{c}_{\un ,}{}^{n(k)}$ is exact for
all $k\neq 1$. Since $y_{\un\um}$ is not required to be traceless
there is a nontrivial cohomology class
\be
\hat{c}_{\un ,}{}^n = \ga e_{\un ,}{}^n
\ee
with an arbitrary parameter $\ga$.
It is obvious that this element is \sgm - closed but not
\sgm - exact because
\be
 e_{\un ,}{}^n\neq e_{\un ,m} X^{nm}
\ee
for any traceless $X^{nm}$. {}From the analysis of the sections 2 and 3
it is
clear that sending this cohomology to zero in (\ref{cc}) is equivalent
to imposing the Klein-Gordon equation.  The fact that all other
\sgm - cohomology groups in $H^1$ vanish means that the rest equations
in the chain (\ref{chain}) and it's $AdS$ generalization
contain no further differential restrictions on the field $c(x)$,
merely expressing
higher components $c^{n(k)}$ in terms of derivatives of $c$.

To summarize, the equation (\ref{cald0})
contains one differential equation, Klein-Gordon equation,
along with an infinite set of
constrains expressing the fields $c^{n(k)}$ $k\geq 1$ via
derivatives of the physical field $c$ according to
(\ref{KG}) and (\ref{hder}) in the
flat space or (\ref{adsKG}) and (\ref{adshder})
in $AdS_d$.

Note that the dynamical field $c$ represents a nontrivial cohomology group
$H^0$. Indeed, $c$ is
\sgm - closed but, being a 0-form, cannot be \sgm -exact.
Let us stress
that the physical field $c$ should be \sgm -closed to make
it possible to start the analysis of the equations by
writing (\ref{start}).

One of the lessons of this section is that,
in order to have a chain of equations $\tilde{\cal D}\tilde C =0$
 which does not impose differential
restrictions on a bottom field(s) merely expressing
some of the 0-forms $\tilde{C}$ in
terms of derivatives of some other, one has to modify the
setting in such a way that  $\tilde H^1$=0.
Such a problem setting is expected to be
useful at the nonlinear level
for the off-mass-shell formulation of the action
principle compatible with
the constraints written
in the covariant form $\tilde{\cal D}\tilde C = 0$.
For a free scalar field
is developed in the section \ref{Off-Mass-Shell System}.

\section{Free Action}
\label{Free Action}

In \cite{V2,LV} it has been shown that the ``higher spin''
covariant derivatives analogous to (\ref{adscovd}) can be used to build
action functionals. Such a form of the higher spin action
was used in \cite{FV1} to introduce higher-spin-gravitational
interactions at the cubic level.
In this section we show how the free action for
a scalar field can be formulated in terms of the
covariant derivative (\ref{adscovd}). This action is expected to
result from the linearization of some full higher spin  invariant
action describing higher spin and lower spin fields and formulated
in terms of higher spin covariant derivatives.
Hopefully, the proposed action will help to shed some light on the
structure of a nonlinear higher spin action in any dimension
and, in particular, to build the full nonlinear action in $d=4$
containing necessary lower spin matter fields.

We identify the fields $c$ and $c^n$ with the fields
of the first-order Klein-Gordon action. Let us
address the question whether there exists
an action of the form
\begin{equation}
S^2=\sum_{k=1}^\infty
\frac{g(k)}{k!}\int_{M_d}e^{m_{3}}\f\cdots \f e^{m_{d}}\f
\epsilon_{m_{1}\ldots m_{d}}
(\D C)^{m_1 n(k-1)}
\f (\D C)^{m_2} {}_{n(k-1)}
\,, \label{ac1}
\end{equation}
with some coefficients  $g(k)$
(no symmetrization with respect to the indices $m$)
that the equations of motion
derived from this action are equivalent to the spin 0
equations in the form
\bee
(-D_n c^n +m^2)c=0\,,\label{equa1} \qquad c^n=-D^n c\,.
\label{equa2}
\eee
 In addition we require that
\begin{equation}
\label{ac_}
\frac{\delta S^2}{\delta c^{n(k)}}\equiv 0
\mbox{\quad for }k\geq 2\,.
\end{equation}
The latter condition admits a natural
interpretation in view of the formula (\ref{adshder}) as
the requirement that
the free action does not contain higher derivatives of the scalar
field $c$.

To simplify formulae, let us introduce the operator
$E\mbox{:  }T^{p}_{k}\rightarrow T^{d+p-2}_{k}$
and
$N\mbox{:  }T^{p}_{k}\rightarrow T^{p}_{k}$,
\be
(EC)^{n(k)}=g(k)e^{m_{1}}\f\cdots \f e^{m_{d-2}}\f c^{n(k-1)t}
\epsilon_t{}^n{}_{m_{1}\ldots m_{d-2}}\,,\quad k\geq 1\,; \qquad
E(c)=0\,,
\label{e}
\ee
\be
(NC)^{n(k)} =kc^{n(k)}\,
\ee
for any p-form $c^{n(k)}$. The following
elementary identities take place
\begin{equation} \sgp
N=(N-1)\sgp \,,\qquad \sgm N=(N+1)\sgm\,,
\end{equation}
\begin{equation}
E\sgm=(-1)^{d}\frac{N+1}{N+d-1}\frac{g(N)}{g(N+1)}\sgm E \,,\quad N\geq 1\,,
\label{c1}
\end{equation}
\begin{equation}
E\sgp=(-1)^{d}\frac{N+d-2}{N}\frac{g(N)}{g(N-1)}\sgp E   \,,\quad N\geq 2\,.
\label{c2}
\end{equation}

Let us introduce notation
\be
\label{form}
(C^p , B^q ) = \sum_k
 \frac{1}{k!}c^{n(k)}\f
b_{n(k)}\,,
\ee
where $C^p \in T^p$ and  $B^q \in T^q$.
The following identities are true
\begin{equation}
\al C^{p},B^{q}\ar=(-1)^{pq}\al
B^{q},C^{p}\ar\;,
\end{equation}
\begin{equation}
\al C^{p},EB^{q}\ar=(-1)^{pd-1}\al
EC^{p},B^{q}\ar\,,
\end{equation}
\begin{equation}
\al C^{p},\sgp B^{q}\ar=
(-1)^{p}\al \frac{f(N)}{N+1}
\sgm C^{p},B^{q}\ar\,,
\label{p1}\end{equation}
\begin{equation}
\al C^{p},\sgm B^{q}\ar=
(-1)^{p}\al\frac{N}{f(N-1)}
\sgp C^{p},B^{q}\ar\,.
\label{p2}
\end{equation}

With this notation, the action (\ref{ac1}) reads
\begin{equation}
 S^2=\int_{M_d} \al E\D C,\D C\ar .
\label{ac2}
\end{equation}
Since the form (\ref{form}) is Lorentz invariant, the
Stokes theorem can be written in the form
\begin{equation}
\int_{M_d}\al C^{p},DB^{d-p-1}\ar=-(-1)^{p}\int_{M_d}
\al DC^{p},B^{d-p-1}\ar+(-1)^p\int_{\partial M_d}
\al C^p,B^{d-p-1}\ar\,.
\label{D}
\end{equation}

A local variation of the action is
\be
\delta S^2= \int_{M_d}\al E\D C,\D\delta C \ar+
\int_{M_d}\al E\D\delta C,\D C\ar=
2\int_{M_d}\al E\D C,\D\delta C \ar \,.
\ee
Integrating by parts and using the definition (\ref{adscovd})
of the operator \D{} along with the
identities (\ref{c1}), (\ref{c2}), (\ref{p1}) and (\ref{p2})
we obtain
\begin{eqnarray}
\delta S^2=(-1)^{d-1}2\int_{M_d}\left (\left[\left[\frac{N+d-2}{N}
\frac{g(N)}{g(N-1)}+\frac{N}{f(N-1)}\right]\sgp +\right.\right. \nonumber\\
\left.\left.+\left[\frac{N+1}{N+d-1}
\frac{g(N)}{g(N+1)}+\frac{f(N)}{N+1}\right]\sgm
\right]E\D C,\delta C\right )\,,
\label{varac}
\eee
provided that $\delta C=N(N-1)\Upsilon$ for arbitrary $\Upsilon$
(i.e. $\delta C$ does not contain $\delta c$ and $\delta c^n$).
The condition that this variation vanishes is equivalent
to (\ref{ac_}).
Requiring the coefficients in front of
\sgm{} and \sgp{} to vanish we obtain the set of equations
on $g(k)$
\be
g(k)=-\frac{g(k-1)}{f(k-1)}\frac{k^2}{k+d-2}\,,
\label{gkeq}
\ee
which admits a unique solution up to an arbitrary normalization
factor that can be fixed as
\be
g(1)=\frac{1}{2}\frac{d^2}{m^2}\,,
\ee
to have  the standard normalization of the free scalar field
action.

Substituting $f(k)$ (\ref{adsf_}) we obtain for the case of
an arbitrary mass\footnote{For the values of
mass $m^2=\lambda^2k_0(k_0+d-1)$, $k_0=0,1,\ldots$
the function $g(k)$ tends to infinity when $k>k_0$ thus
making the formula for the action (\ref{ac1}) inapplicable.
These special values of the parameter of mass are analogous
to those found previously in \cite{BPV} at the level of
equations of motion for $d=3$. Their appearance
signals that for these special values of the parameter the infinite
chains start not with the scalar $c$, but rather with the rank
$k_0$ tensor $c^{n(k_0)}$ so that after
appropriate redefinition of the normalization constant
the action (\ref{ac1}) turns out to be well-defined. We hope to
consider this interesting question elsewhere.}
in $AdS_d$ space
\be
g(k)=\frac{(-1)^{k-1}}{2}\frac{1}{m^2\lambda^{2k-2}}
\frac{\Gamma(\ga^++2)
\Gamma(\ga^-+2)}{\Gamma(k+\ga^++1)\Gamma(k+\ga^-+1)}
\frac{d!}{(d-2)!!}\frac{(2k+d-2)!!}{(k+d-2)!}k!\,, \ee
where
\be
\ga^\pm=\frac{1}{2}\Big(d-3\pm \sqrt{(d-1)^2+4\frac{m^2}{\lambda^2}} \Big)\,.
\ee
For the flat case with nonzero mass we have

\be
g(k)=\frac{(-1)^{k-1}}{2}
\frac{1}{m^{2k}}\frac{d!}{(d-2)!!}\frac{(2k+d-2)!!}{(k+d-2)!}k!\,.
\ee

Note that $g(k)$ is such that, by virtue of (\ref{c1}) and (\ref{p1}),
\sgm{} is conjugated to \sgp{}
with respect to the form $\al E(C) ,B \ar$ restricted to $T_k$ with
$k\geq 2$, i.e.
\begin{equation}
\al EC^{p},\sgp B^{q}\ar=(-1)^{p-1}\al
E\sgm C^{p},B^{q}\ar\;\;\;\mbox{for }C^p=N(N-1)\Upsilon^p\,.
\label{ip}
\end{equation}
With the help of this property it is trivial to see that $\delta S^2=0$
provided that $N(N-1)\delta C =0$.
This explains why we have obtained only one equation (\ref{gkeq}) from the
requirement that the two coefficients vanish in (\ref{varac}).

Since the
 conjugation relation (\ref{ip}) does not take place for $T^0_0$ and
$T_1^0$,  the equations of motion for the fields $c$ and
$c^n$ turn out to be nontrivial having a form
\be
\frac{\delta S^2}{\delta c}=2(-1)^d
\frac{m^2}{d}\sgm E(\D C|_{T_1^1})=0\,,
\label{eqm1}
\ee
\be
\frac{\delta S^2}{\delta c^1}=-2E\sgp (\D C|_{T^1_0})=0\,.
\label{eqm2}
\ee

It is elementary to see that these equations are equivalent to
the free field equations (\ref{equa1}). In fact, the equation
(\ref{eqm2}) is equivalent to
\be
(\D C)|_{T_0^1}=0\,
\ee
(i.e., ${\rm Ker} (E \sgp )|_{T^1_0} =0 $), while the equation (\ref{eqm1})
is equivalent to
\bee
&(\D C)|_{T_1^1}=\sgm X
\eee
with an arbitrary $X$
(${\rm Ker} (\sgm E )|_{T_1^1}=\sgm X $). One can always adjust such a field
$c^{nm}$  that $X=0$. This condition
can be interpreted as some
constraint on $c^{nm}$. As shown in the sections 2, 3 and 4, by imposing
constraints on the higher fields $c^{n(k)}$ one can achieve $\D C=0$
provided that the dynamical equations (\ref{eqm1}) and (\ref{eqm2})
are satisfied.

Although the bilinear action is independent of (local variations of)
the ``extra fields" $c^{n(k)}$ with
$k\geq 2$,
it is useful to express ``extra fields" $c^{n(k)}$ with $k\geq 2$
in terms of the
derivatives of the dynamical scalar field $c$ because extra fields
contribute at the interaction level and have to be taken
into account in the boundary terms. The natural choice for these
constraints is according to (\ref{adshder}).
At the action level
this can be achieved by adding the
term

\be
S^c=\sum_{k=2}\int_\Omega \gamma^{n_1\ldots n_k}(c_{n_1\ldots n_k}-(-1)^k
D_{n_1}\cdots D_{n_k}c)\,,
\label{sc}
\ee
with Lagrange multipliers
 $\Gamma=(\gamma^{n(2)},\gamma^{n(3)}\ldots)$,
($\gamma^{n(k)}$ is symmetric and traceless).
The action
\be
S=S^2 +S^c
\label{action}
\ee
leads to the equations (\ref{eqm1}) and (\ref{eqm2}) along with
\bee
\label{cons}
c_{n_1\ldots n_k}=(-1)^k\p D_{n_1}\cdots D_{n_k}c\,,\qquad
\gamma^{n(k)}=0\,,\quad
k\geq 2\;.
\eee
Taking into account the results of the section \ref{sec_cog},
these equations are equivalent to
\begin{eqnarray}
\D C&=&0 \,, \label{eqdcz}\\
\Gamma&=&0\label{eqlz}\,.
\end{eqnarray}

Let us note that it is possible to extend the set of Lagrangian
multipliers by adding $\gamma^n$ and $\gamma$
The scalar $\gamma$ cancels out from the action.
The term with $\gamma^n$ contributes but leads to
the equivalent dynamics with $\gamma^n =0$ as a consequence
of the equations of motion.

So far, we have considered  local variations in the bulk.
If  $\partial M_d$ is nontrivial, the boundary terms
have to be taken into account.
It is elementary to see directly that
\bee
&S^2=
\int_{M_d}\Big(2\al E\sgp c,Dc^n\ar +\al E\sgp c,\sgp c\ar
-\al E\sgp \sgm c^n,c^n\ar\Big)+\nn\\
&{}+(-1)^{d-1}\int_{\partial M_d}
\Big( \al E\D C,C\ar -\al E \sgp c,c^n\ar \Big)
\,.
\label{excac}
\eee
Thus, the action $S^2$ equals to the standard first-order Klein-Gordon
action modulo boundary terms which have to be subtracted to make the
two actions equivalent.

\section{Fock Space Notation}
\label{Fock Space Notation}

Instead of working with infinite sets of tensors $C$ it is
convenient to use the Fock-space  language analogous to
that used in \cite{LV} for spin $s\geq 1$ fields. Namely, we introduce
the creation and annihilation operators  \ah{n}{} and \ac{m}
obeying the  commutation relations
\be
[\ah{n},\ac{m}]= \eta_{nm}\,,
\ee
\be
[\ah{n},\ah{m}]= [\ac{n},\ac{m}]=0\,.
\ee
Given set of p-forms $c^{n(k)}$, we introduce a Fock vector
\be
\label{fs}
|c(p) \cet=\sum_{k=0}^\infty |c(p,k)\cet=
\sum_{k=0}^\infty \frac{1}{k!}a^+_{n_1}\cdots a^+_{n_k}
c^{n_1 \ldots n_k}|0\cet\,.
\ee

It is convenient to introduce operators
\be
N^{++}=\half a^+_n a^{+n}\,,\qquad
N^{--}=\half a_n a^{n}
\ee
and
\be
N=a^+_n a^n \,,
\ee
satisfying the        commutation relations
\be
[N,N^{++} ]=2N^{++}\,,\qquad
[N,N^{--} ]=-2N^{--}\,,
\ee
\be
\label{sl2}
[N^{--} ,N^{++} ]=N+\frac{d}{2}\,,
\ee
which transform to the $sl_2$ commutation relations
by the trivial shift $N^0 =N+\frac{d}{2}$.

The subspace of
traceless symmetric tensors is extracted by the condition
\be
\label{N--0}
N^{--}|c(p,k)\cet=0\,.
\ee
The explicit form of the projector $\p$ to the traceless tensors is
complicated. For practical computations it is however enough to use
the following simple formulae
\be
\label{pr1}
a^+_n  \p = \p a^+_n + (N+\frac{d}{2} -2)^{-1} N^{++} a_n \p \,,
\ee
\be
\label{pr2}
a_n  \p = \p a_n + (N+\frac{d}{2} -1)^{-1}\p a^+_n  N^{--} \,.
\ee

We have
\be
\sgm=e_n a^n\,,
\ee
\be
\label{sg+f}
\sgp=\frac{f(N-1)}{N}\p e^na^+_n =
\frac{f(N-1)}{N}\left(e^na^+_n-(N+\frac{d}{2} -2)^{-1}
N^{++}\sgm \right)\,,
\ee
\be
D^n=e^{\un n}(\partial_\un+\omega_\un{}^t{}_m a^+_t a^m)\,,
\ee
\be
E=\frac{g(N)}{N}e^{t_1}\f\cdots\f e^{t_{d-2}}
\epsilon_m{}^n{}_{t_1\ldots t_{d-2}}a^+_n a^m \,.
\ee

It is also useful to use   Lorentz invariant ``base'' oscillators
\be
a_\un = e_\un{}^n a_n\,,\qquad a^+_\un = e_\un{}^n a^+_n\,,\qquad
a^\un = e^\un{}^n a_n\,,\qquad a^{+\un} = e^{\un n} a^+_n\,,
\ee
satisfying
\be
\label{Da}
D (a_\un )=0\,,\qquad D (a^\un )=0\,,\qquad
D (a^+_\un )=0\,,\qquad D (a^{+\un} )=0\,
\ee
by virtue of the zero torsion condition (\ref{Ra}).
Using the definition (\ref{fleq}) of ${\cal D}$ and introducing
\be
D^+ = D^n a_n^+ = D_\un a^{+\un}
\ee
we rewrite the action (\ref{action}) in the form
\be
S=(-1)^{d-1}\langle\D C|E|\D C\rangle+\langle
\Gamma | C\ra -\la \Gamma |
{\rm exp}(-D^+)c|0\rangle \,.
\ee

\section{Yang-Mills Interaction}

Although the free action (\ref{ac2}) was shown to be equivalent
modulo some boundary terms to the
usual first-order Klein-Gordon action, the
minimal Yang-Mills interaction introduced
via covariant derivatives leads to different results in the two
cases. This is not surprising because in the equivalence
proof we have used the fact that ${\cal D}^2=0$
for the background $AdS$ geometry. This is no longer true
in the presence of the Yang-Mills connection $A$
\be
{\cal D} \rightarrow {\cal D}^{YM} = {\cal D} +A\,,
\ee
with
\be
\label{DDF}
({\cal D}^{YM} )^2=G\,,\qquad G=dA +A\wedge A\,.
\ee
Here $A$ and $G$ take values in some representation $t$ of the gauge
group $g$, in which the scalar fields $c^{n(k)}$ take their values
(i.e. $c^{n(k)}\rightarrow c^\ga {}_,{}^{n(k)}$,
$A\rightarrow A^\ga{}_\gb$).
The two actions can therefore differ by some terms proportional
to the Yang-Mills field strength $G$.

An interesting consequence of the reformulation of the scalar
field action in the form (\ref{ac2}) is that it immediately leads to
 the pseudolocally exact form of the conserved spin-1 current
generalizing the d=3 results of \cite{PV2} for spin -1 currents to
the case of  scalar matter field of arbitrary mass in $AdS_d $ or
nonzero mass in the flat space. Indeed, the action (\ref{ac2})
with the minimal Yang-Mills interaction reads
\be
\label{actym}
S^{gau}=\int_{M_d} tr\al E\D^{YM} C,\D^{YM} C\ar \,.
\ee
The spin-1 conserved current is
\be
J^{\un}{}_{ \ga}{}^\gb=\frac{\delta}{\delta A_{\un}{}^{\ga}{}_\gb}
S^{gau} \mid_{A=0}\,.
\ee
For the action (\ref{actym}) we have
\be
\delta S^{gau}|_{A=0}=
\int_{M_d}\Big(
\al E\delta A(C),\D C\ar +\al E\D C,\delta A(C)\ar\Big)
=2\int_{M_d}\al E\D C,\delta A(C)\ar
\ee
Taking into account that the free equations of motion (\ref{eqm1}),
(\ref{eqm2})
along with the constraints (\ref{cons}) imply $\D C=0$ we arrive
 at the paradoxical result that the
Yang-Mills current derived from the action (\ref{ac2}) vanishes
on-mass-shell despite the fact that the action is explicitly
invariant under the Yang-Mills symmetry\footnote
{Let us note that since the variation of the free action with respect
to the extra fields vanishes identically, it is enough for
the analysis of the cubic interaction to use the free constraints
(\ref{adshder}). In other words, corrections due to the Yang-Mills
covariantization of the expression (\ref{sc}) may only affect quartic
interactions. Note that the term with Lagrange multipliers does
not affect this consideration either because $\Gamma=0$ on-mass-shell.
Alternatively, one can impose the constraints by hand without
introducing Lagrange multipliers.}.
An important  related comment is that the proposed formulation is
applicable just in those cases when either the $S$-matrix cannot
be defined ($AdS$ case) or the contribution of the corresponding
three-particle vertices to the scattering amplitude vanishes
by kinematical reasons (flat space case with $m\neq 0$).

Usually one argues that
any terms in the current interaction that vanish on-mass-shell
are irrelevant because one can compensate them by some local field
redefinition
\be
\label{fr}
C\to C^\prime=C+AC^2\,.
\ee
 In the case under consideration
one has to take into account however that the variation contains
infinitely many terms with all derivatives of the scalar
field $c$ via the highest components $c^{n(k)}$ (\ref{adshder}).
Therefore, a field redefinition (\ref{fr})
compensating interactions in the action (\ref{ac2})  may be
nonlocal. Such expressions containing infinite series
in powers of derivatives  were called in \cite{PV2}  pseudolocal.  It is
not surprising that some interaction can be compensated by a nonlocal
field redefinition (for example with the aid of the Green function).
The naive conclusion that the action (\ref{actym}) does not describe
nontrivial interactions is therefore wrong.

The lesson is that usual current interactions playing a fundamental
role in the local field theory  may not be that important
in theories with interactions containing infinite series in
higher derivatives. This is true both for massive modes in the
flat space (like in the string theory) and for theories with arbitrary
mass in the $AdS$ background like in the higher spin theories.
One way to see this
is to show that ordinary Yang-Mills current is on-mass-shell exact
in the class of pseudolocal expansions. To this end, let us
compare the action (\ref{ac2}) with the standard
first-order Klein-Gordon action with the Yang-Mills interaction.
Proceeding as in the section \ref{Free Action}
we arrive at the following result
\bee
&S^{gau}=
&\int_{M_d}\Big(2\al E\sgp c,D^{YM} c^n\ar +\al E\sgp c,\sgp c\ar
-\al E\sgp\sgm c^n,c^n\ar\Big)+
\int_{M_d}\al EGC,C\ar\nn\\
&&{}+(-1)^{d-1}\int_{\partial M_d}\Big(
\al A\D^{YM} C,C\ar-\al E \sgp c,c^n\ar\Big)\,,
\label{covaction}
\eee
where $D^{YM}=D+A$.
The first term in this formula just describes the covariantized
Klein-Gordon first-order action $S_{KG}^{gau}$.
Therefore, we obtain that the difference
\be
\label{dss}
\Delta = S^{gau} - S_{KG}^{gau}
\ee
is proportional to the Yang-Mills field strength up to
some boundary terms
\be
\Delta=\int_{M_d}\al EGC,C\ar+\mbox{boundary terms} \,.
\label{delta}
\ee
{}From this formula one immediately derives
the pseudolocally exact representation
for the spin-1 current. Indeed,
\be
S_{KG}^{gau}=S_{KG}^{free}+ \int_{M_d}tr( {}^*J\wedge A ) +O(A^2 )\,.
\ee
On the other hand we have
\be
\frac{\delta S^{gau}}{\delta A_{\un}{}^{ \ga}{}_\gb } |_{A=0} \sim 0\,,
\ee
where the $\sim$ implies ``on-mass-shell" equality and
\be
\Delta= \int_{M_d}tr(U\wedge G  )\,,\qquad
U=\al EC,C\ar \,.
\label{deltau}
\ee
 Taking into account (\ref{dss}) we obtain

\be
{}^*J\sim (-1)^d dU
\ee
with $U$ (\ref{deltau}) being an infinite expansion in
higher derivatives due to (\ref{cons}).

An interesting problem for the future is to
study the role of the boundary terms in (\ref{covaction})
in the context of the AdS/CFT correspondence \cite{adsconf}
to clarify to which extend
the dynamics of the bulk action (\ref{covaction}) is encoded in
the  boundary actions.

\section{Off-Mass-Shell System}
\label{Off-Mass-Shell System}

The approach developed in this paper is analogous to that developed
in \cite{LV} for massless gauge fields in any dimensions and
in \cite{Ann} for all massless fields in d=4. It is  adequate
for the analysis of the nonlinear
equations of motion as some deformation of (\ref{eqdcz}).
Also it is useful for the analysis of the interactions
of higher spin gauge fields
at the cubic level because the analysis of the Noether current
interactions is essentially on-mass-shell\footnote{Note that
the situation with higher spin gauge fields
considered in \cite{FV1} is different from the one with the scalar
field discussed in this paper in that respect that,
for any three fixed spins,  only a finite
number of terms that vanish on-mass-shell appear in the gauge
variation of the higher spin action of Ref.\cite{FV1}
and, therefore, the fact that there exists some deformed gauge
transformation that leaves the action invariant \cite{FV1} is the
well-defined local statement.} so that constraints
and equations of motion can be used simultaneously in the form
analogous to (\ref{eqdcz}). Beyond the cubic level
 one needs however appropriately modified off-mass-shell constraints
compatible with the higher spin symmetries. In \cite{V2} the
problem of formulation of invariant constraints was called the
``extra field" problem.  Here we undertake a
step in this direction solving the problem of separation of
constraints and dynamical field equations
for the simplest case of a free scalar field.

The idea is to generalize the
 approach of the section \ref{sec_cog} in such a way that
the generalized covariant constancy conditions
\be
\label{cc_}
\Dg \ti C=0
\ee
for some extended set of the fields $\ti C$ express all the fields
in terms of derivatives of the dynamical scalar field $c (x)$
imposing no differential restrictions on the latter.
As shown in the section \ref{sec_cog}, nontrivial differential equations on
the physical field $c$ are associated with the cohomology of
the operator \sgm . The idea therefore is to look for such a
modified covariant derivative $\Dg \ti C$ ($\Dg^2 = 0$)
that the \sgmg{}-cohomology of \Dg{} is trivial. An additional requirement
is that together with the Klein-Gordon equation for $c$, the equation
(\ref{cc_}) should be equivalent to the system (\ref{adsKG}),
(\ref{adshder}).
These conditions  can be achieved by extending
the set of tensors $C$ to all symmetric but not necessarily
traceless tensors $\ti C=(\ti c,\ldots,\ti c^{n(k)},\ldots)$.

Let $\ti T^p_k$ be
a linear space of $p$-forms taking values in the
space of rank-$k$ totally symmetric tensors
and $\ti T^p = \sum_{k=0}^\infty \oplus \ti T^p_k$,
$\ti T = \sum_{p=0}^\infty \oplus \ti T^p$.
The space $\ti T^p $ can be realized as a Fock space
of the section \ref{Fock Space Notation}  relaxing the
tracelessness  condition (\ref{N--0}).

Let the operator \sgmg  $\ti T^p_k\rightarrow \ti T^{p+1}_{k-1}$
be defined as before
\be
\sgmg = e^n a_n \,
\ee
(equivalently, in terms of components, $(\sgmg \ti C)^{n(k-1)}=
e_{m}\f \ti c^{n(k-1)m}).$
Obviously,
\bee
&\sgmg\sgmg=0\label{sgmg1}\,,\\
&D\sgmg+\sgmg D=0
\label{sgmg2q}\,.
\eee
Note that the operator \sgmg{} is different from \sgm{}
because it acts in a different space. This is manifested by the
fact that, as is necessary for our construction,
the first \sgmg{}-cohomology group $\ti H^1$ is trivial.
(The explicit proof of this fact is not given here because
it is a simplified version
of that for the \sgm-cohomology in the section \ref{sec_cog}.)
On the other hand, \sgmg{} leaves invariant the subspace $T \in \ti T$
spanned by traceless tensors and \sgm{} is the restriction of \sgmg{} to $T$
\be
\sgmg|_{T}=\sgm \,.
\label{sgmg4}
\ee

Let us look for the operator $\Dg$ in the form
\be
\Dg= D+\sgmg+\sgpg \,,
\label{gsy}
\ee
where \sgpg: $\ti T^p_k\rightarrow \ti T^{p+1}_{k+1}$
is some operator demanded to satisfy the conditions
\bee
&\sgpg\sgpg=0\label{sgpg1}\,,\\
&D\sgpg+\sgpg D=0
\label{sgpg2}\,,\\
&\{\sgpg,\sgmg\}=-DD=
-\lambda^2e^{n}a^+_n  e^m a_m
\label{sgpg3} \,
\eee
to guarantee the
compatibility condition
\be
\Dg\Dg=0\,.
\ee
In addition it is convenient to require that
\be
\sgpg|_{T}=\sgp
\label{sgpg4}\,
\ee
to interpret the on-mass-shell chain
$\D C=0$ as the restriction of (\ref{cc_}) to $T$.

Let us look for the operator \sgpg{} in the form
\be
\sgpg = p(N) e^n a^+_n + q(N) N^{++} e^n a_n
\label{x}
\ee
with some coefficients $p(N)$ and $q(N)$. The conditions
(\ref{sgpg3}) and (\ref{sgpg1}) give rise to the following equations
\be
q(N+1) = p(N+1) -p(N) +\lambda^2
\ee
and
\be
p(N) q(N-1) - q(N)p(N-1) +q(N) q(N-1) =0 \,,
\ee
respectively. The condition  (\ref{sgpg4}) is equivalent to the
requirement that $N^{--} \sgpg = X_+ N^{--}$ with some operator $X_+$.
By virtue of (\ref{sl2}) it gives
\be
p(N) = -(N+\frac{d}{2}-2) q(N)\,.
\ee
The generic solution of all these conditions reads
\be
p(N)=\half\left( \lambda^2 (N+\frac{d}{2} -2 )
+\frac{m_c^2}{N+\frac{d}{2} -1}\right)\,,
\ee
where $m_c^2$ is an arbitrary parameter. When \sgpg{} is applied
to a traceless
vector it reproduces  the operator \sgp (\ref{sg+f}), (\ref{adsf})
of the on-mass-shell problem with
\be
m_c^2 =m^2+ \frac{\lambda^2}{4}d(d-2)\,.
\ee
Note that $m_c^2 =0$ corresponds to the conformal case
(see footnote 3).

To summarize, we have shown that on-mass-shell covariant derivative
(\ref{adscovd}) admits such a  generalization to a larger set
of fields that the covariant constancy conditions (\ref{cc_})
do not impose any dynamical equations on the matter field $c(x)$
merely expressing higher components in the set ${\ti C}$ via higher
derivatives of $c(x)$. Because the operator $ \Dg$ is
defined in such a way that it reduces to $\D$ when restricted to
the subspace $T^0\subset \ti T^0 $,
the dynamical field equations turn out to be equivalent to the
condition that all fields in ${\ti T^0}/T^0$ vanish. It is an
interesting problem for the future to find an action principle
leading to such field equations.

\section{Conclusions}

In this paper the dynamics of a scalar field in $AdS_d$
is formulated in terms of certain ``higher spin'' covariant derivatives
both at the level of equations of motion and at the Lagrangian level.

Interestingly enough the proposed formalism leads to the
interpretation of the dynamical field  equations (i.e. Klein-Gordon
equation) as the requirement that the fields belong to the
trivial class of certain cohomology group, \sgm-cohomology.
An interesting problem for the future is to extend this
interpretation to other types of relativistic fields and to
clarify its group-theoretical meaning.

The new action principle for a scalar
field in arbitrary dimension proposed in this paper is shown
to be equivalent (modulo boundary
terms) to the standard first-order Klein-Gordon action
at the free field level but
 different at the interaction level leading to pseudolocal
interactions containing derivatives of all orders. This action is
defined for massive fields in the flat space and for fields of
an arbitrary mass
(requiring some redefinition for special values of
$\frac{m^2}{\lambda^2}$ --- see footnote 4)
in $AdS_d$. It contains inverse powers of
either the  parameter of mass or the cosmological constant in front of
the terms with higher derivatives. In this sense it is analogous to
the higher spin actions constructed previously in \cite{V2} and to
the actions for massive higher spin fields constructed recently in
\cite{per_,per}. This picture fits nicely the superstring
picture in which interactions contain powers of the parameter
$\ga^\prime$ that fixes (inverse) mass scale in the theory.

An interesting conclusion of this paper is that
current interactions do not play a fundamental role in the theories
admitting infinite expansions in higher derivatives at the interaction
level like higher spin theories and string theories. This conclusion
is in fact welcome for any theory expected to be
identified with one or another phase of the
string theory
because ordinary local field theory
Feinman diagram expansion (i.e., with local current vertices)
contradicts
duality in the string theory \cite{GSW}. The scalar field action
presented in this paper illustrates how the ordinary field-theoretical
actions reformulated in the higher spin inspired way can escape this
potential conflict.
An important  related point is that the proposed formulation is
applicable just in those cases when either the $S$-matrix cannot
be defined ($AdS$ case) or the contribution of the corresponding
three-particle vertices to the scattering amplitude vanishes
by kinematical reasons (flat space case with $m\neq 0$).

\section*{}

The authors are grateful to S.Prokushkin for collaboration at the
early stage of the work and also to R.Metsaev  and
I.Tipunin for useful discussions.
This research was supported in part by
INTAS, Grant No.96-0538 and by the RFBR Grant No.99-02-16207.

\end{document}